\documentclass[twocolumn,showpacs,preprintnumbers,amsmath,amssymb,prl,superscriptaddress,a4paper,dvipdfmx]{revtex4}
\usepackage{graphicx,here}% Include figure files
\usepackage{dcolumn}% Align table columns on decimal point
\usepackage{bm}% bold math
\usepackage{enumerate}
\usepackage{graphicx}

\usepackage{ulem,color}

\newcommand{\lag}{\langle}
\newcommand{\rag}{\rangle}
\newcommand{\lt}{\left}
\newcommand{\rt}{\right}

\newcommand{\mrm}{\mathrm}

\newcommand{\mcl}{\mathcal}

\begin{document}

%\preprint{APS/123-QED}
\date{\today}

\title{Cluster-based Haldane state in edge-shared tetrahedral spin-cluster chain:\\ Fedotovite K$_2$Cu$_3$O(SO$_4$)$_3$}

\author{M. Fujihala}
\email{fujihara@nsmsmac4.ph.kagu.tus.ac.jp}
\affiliation{
Department of Physics, Faculty of Science, Tokyo University of Science, Shinjuku, Tokyo 162-8601, Japan
}%
\author{T. Sugimoto}
\email{sugimoto.takanori@rs.tus.ac.jp}
\affiliation{
Department of Applied Physics, Faculty of Science, Tokyo University of Science, Katsushika, Tokyo 125-8585, Japan
}%
\author{T. Tohyama}
\affiliation{
Department of Applied Physics, Faculty of Science, Tokyo University of Science, Katsushika, Tokyo 125-8585, Japan
}%
\author{S. Mitsuda}
\affiliation{
Department of Physics, Faculty of Science, Tokyo University of Science, Shinjuku, Tokyo 162-8601, Japan
}%
\author{R. A. Mole} 
\affiliation{
Australian Nuclear Science and Technology Organization, Lucas Heights, New South Wales 2232, Australia
}%
\author{D. H. Yu} 
\affiliation{
Australian Nuclear Science and Technology Organization, Lucas Heights, New South Wales 2232, Australia
}%
\author{S. Yano}
\affiliation{
National Synchrotron Radiation Research Center, Neutron Group, Hsinchu 30077, Taiwan
}%
\author{Y. Inagaki}
\affiliation{
Department of Applied Quantum Physics, Faculty of Engineering, Kyushu University, Fukuoka 819-0395, Japan
}%
\author{H. Morodomi}
\affiliation{
Department of Applied Quantum Physics, Faculty of Engineering, Kyushu University, Fukuoka 819-0395, Japan
}%
\author{T. Kawae}
\affiliation{
Department of Applied Quantum Physics, Faculty of Engineering, Kyushu University, Fukuoka 819-0395, Japan
}%
\author{H. Sagayama}
\affiliation{
Institute of Materials Structure Science, High Energy Accelerator Research Organization, Tsukuba, Ibaraki 305-0801, Japan
}%
\author{R. Kumai}
\affiliation{
Institute of Materials Structure Science, High Energy Accelerator Research Organization, Tsukuba, Ibaraki 305-0801, Japan
}%
\author{Y. Murakami}
\affiliation{
Institute of Materials Structure Science, High Energy Accelerator Research Organization, Tsukuba, Ibaraki 305-0801, Japan
}%
\author{K. Tomiyasu}
\affiliation{
Department of Physics, Tohoku University, Sendai 980-8578, Japan
}%
\author{A. Matsuo}
\affiliation{
International MegaGauss Science Laboratory, Institute for Solid State Physics, The University of Tokyo, Kashiwa, Chiba 277-8581, Japan
}%
\author{K. Kindo}
\affiliation{
International MegaGauss Science Laboratory, Institute for Solid State Physics, The University of Tokyo, Kashiwa, Chiba 277-8581, Japan
}%

\begin{abstract}
Fedotovite K$_2$Cu$_3$O(SO$_4$)$_3$ is a candidate of new quantum spin systems, in which the edge-shared tetrahedral (EST) spin-clusters consisting of Cu$^{2+}$ are connected by weak inter-cluster couplings to from one-dimensional array. %
Comprehensive experimental studies by magnetic susceptibility, magnetization, heat capacity, and inelastic neutron scattering measurements reveal the presence of an effective $S$ = 1 Haldane state below $T\cong 4$~K. %
Rigorous theoretical studies provide an insight into the magnetic state of K$_2$Cu$_3$O(SO$_4$)$_3$: an EST cluster makes a triplet in the ground state and one-dimensional chain of the EST induces a cluster-based Haldane state. %
We predict that the cluster-based Haldene state emerges whenever the number of tetrahedra in the EST is $even$. %
\end{abstract}

\maketitle

Quantum spin states in low-dimensional magnets have been extensively studied, because of emergent spin gaps and topological features.  %
In particular, intensive studies of one-dimensional (1D) spin systems have successfully captured the exotic quantum states such as Tomonaga--Luttinger spin-liquid state~\cite{TLL nature_Mat} and Haldane state~\cite{I. Affleck}. %
The magnetic excitations of antiferromagnetic (AFM) chain systems are known to follow the Haldane's conjecture: half-integer spin chains have gapless excitations, whereas spin gaps open in integer spin chains~\cite{haldane's conjecture-1,haldane's conjecture-2}. %
Furthermore, spin gaps open by inter-chain interactions and frustration in half-integer spin systems~\cite{RiceLadder}. %
In fact, the spin gap in $S=\frac{1}{2}$ spin-ladder system IPA-CuCl$_3$ has been observed by neutron scattering, where two spins on a rung couple each other with a ferromagnetic (FM) interactions and thus, the Haldane gap appears~\cite{Matsuda2leg}.
The realisation of such a new effective Haldane chain has been proposed target for a universal quantum gate~\cite{Qbit}.
Therefore, we propose a new direction in Haldane chains based on spin cluster in this Letter. 

The Haldane state is usually constructed by integer spins and a 1D AFM interactions between them.
According to preceding studies on the Haldane state in ladder~\cite{Matsuda2leg,Vekua}, triplet ground states in a local structure can play the role of an effective $S=1$ spin at low temperatures.
The triplet ground states can be generally realized in (i) a spin cluster consisting of $even$ number of $S=\frac{1}{2}$ spins. %
Furthermore,  by connecting them with (ii) a small AFM inter-cluster coupling, an effective Haldane chain is expected at low temperatures. %
To realize such a cluster-based Haldane chain, we study magnetic behaviors of an edge-shared tetrahedral (EST) spin-cluster chain (SCC) K$_2$Cu$_3$O(SO$_4$)$_3$, because its crystal structure seems to satisfy the conditions (i) and (ii).
We actually confirm the cluster-based Haldane chain in this compound based on magnetic susceptibility, magnetization, heat capacity, and inelastic neutron scattering measurements.

\begin{figure}[htb]
\begin{center}
\includegraphics[keepaspectratio,width=8.6cm]{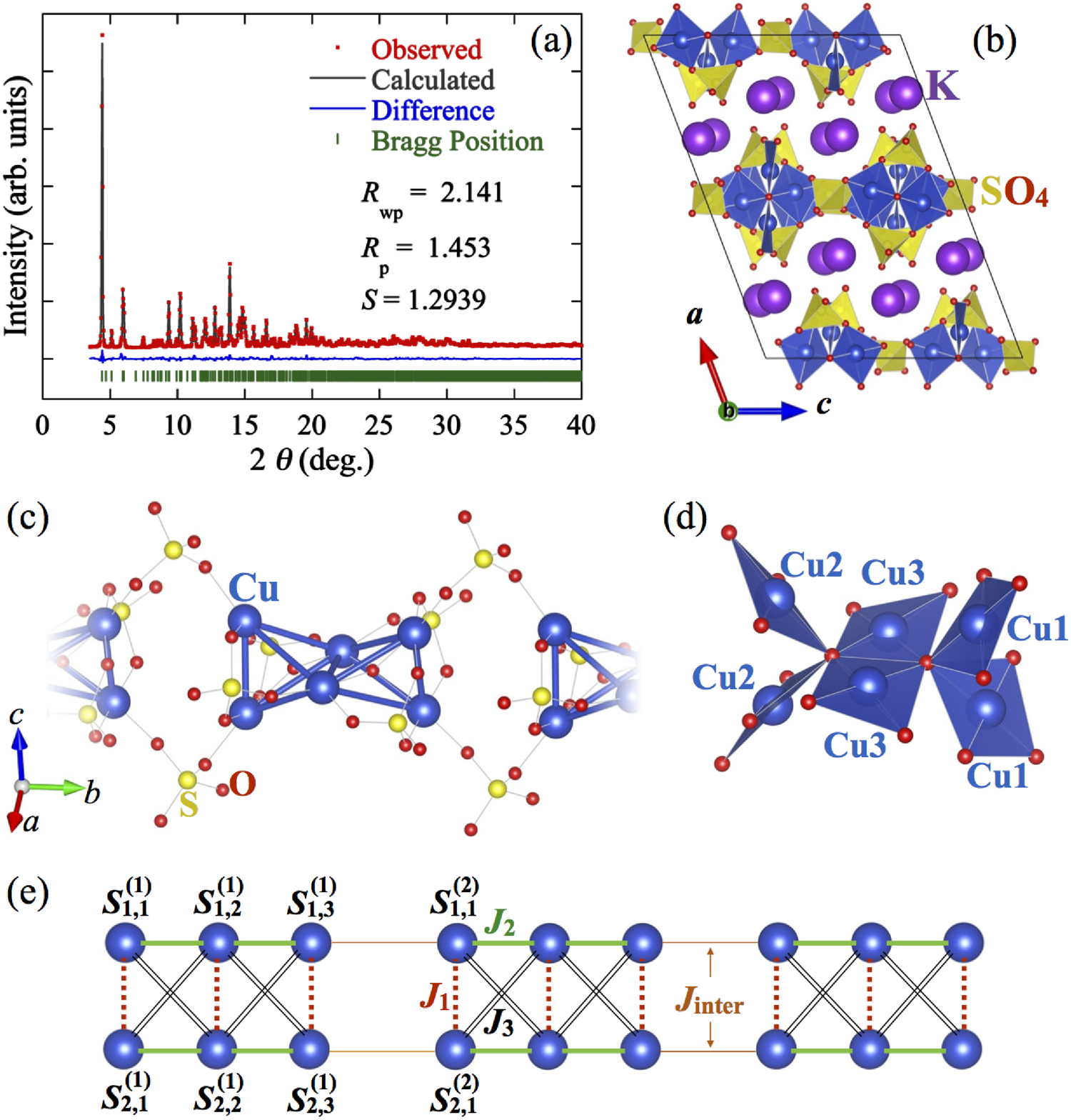}
\caption{(a) Rietveld refinement result of K$_2$Cu$_3$O(SO$_4$)$_3$ using synchrotron XRD data at room temperature.
(b) The crystal structure of K$_2$Cu$_3$O(SO$_4$)$_3$. %
(c) An EST spin-cluster of the Cu$^{2+}$ ions (blue) displayed with nearby oxygen (red) and sulfur (yellow) ions. The EST spin-cluster are connected each other by SO$_4^{2-}$ ions along the $b$-axis. %
(d) Tetracoordinate Cu$^{2+}$ ions and nearby oxygen ions. 
The angles of Cu-O-Cu paths are as follows: $\theta_{\mrm{Cu1-O-Cu1}}=87.0(2)^{\circ}$, $\theta_{\mrm{Cu1-O-Cu3}}=124.2(3)^{\circ}, 110.9(2)^{\circ}$, $\theta_{\mrm{Cu2-O-Cu2}}=90.1(3)^{\circ}$, $\theta_{\mrm{Cu2-O-Cu3}}=125.6(2)^{\circ}, 110.0(3)^{\circ}$, and $\theta_{\mrm{Cu3-O-Cu3}}=98.4(3)^{\circ}, 101.4(3)^{\circ}$.
(e) An effective model for K$_2$Cu$_3$O(SO$_4$)$_3$. Intra-cluster super-exchange interactions $J_1$ (dashed red lines), $J_2$ (green lines) and $J_3$ (black double lines) in an EST spin-cluster, and inter-cluster interaction $J_{\mrm{inter}}$ (orange lines). % 
}
\label{stru}
\end{center}
\end{figure}

The synthesis of K$_2$Cu$_3$O(SO$_4$)$_3$ is designated after the identification of natural mineral Fedotovite~\cite{mineralFed}. %
Single-phase polycrystalline sample is synthesized using a solid state reaction process. %
The space group $C$2/$c$ with lattice parameters of $a$ = 19.0948(6) $\mathrm{\AA}$, $b$ = 9.5282(3) $\mathrm{\AA}$, $c$ = 14.1860(6) $\mathrm{\AA}$, and $\beta$ = 110.644(3)$^{\circ}$ are refined by the computer program RIETAN-FP~\cite{Fujio} (Fig.~\ref{stru}). %
These results are in good agreement with those of the natural mineral Fedotovite~\cite{mineralFedstructure}. %

As illustrated in Figs.~\ref{stru}(b) and (c), magnetic ions of Cu$^{2+}$ form EST spin-clusters, and they are connected each other by SO$_4^{2-}$ ions along the $b$-axis. %
The inter-chain interactions along the $a$- and $c$-axes can be neglected in the experimental temperature range, because the EST-SCCs are separated along the $a$-axis by non-magnetic potassium ions as shown Fig.~\ref{stru}(b), and the tetracoordinated Cu$^{2+}$ ions do not have exchange paths through an anion along the $c$-axis~(see Supplemental Materials Sec.~II~\cite{supple}).
Therefore, we consider that K$_2$Cu$_3$O(SO$_4$)$_3$ has the near-ideal one-dimensional EST-SCCs. %

In the EST cluster, the magnetic couplings originate from super-exchange interactions through the Cu-O-Cu paths (see the caption of Fig.~\ref{stru}).
The super-exchange interaction $J$ in low-dimensional cuprates strongly depends on crystal structural parameters. %
According to Ref.\cite{Misuno}, the magnetic interactions through the Cu1-O-Cu1, Cu2-O-Cu2 and Cu3-O-Cu3 paths are estimated to be small, but the magnitude of the others can be over 10~meV. %
We thus propose the following model Hamiltonian to describe magnetic behaviors of K$_2$Cu$_3$O(SO$_4$)$_3$,
\begin{equation}
 \mathcal{H} = \sum_{k=1}^{L-1} \mathcal{H}_{\mrm{inter}}^{(k:k+1)} + \sum_{k=1}^L \mathcal{H}_{\mrm{intra}}^{(k)}  \label{eq:ham}
\end{equation}
with the inter- and intra-cluster Hamiltonians 
\begin{align}
\mathcal{H}_{\mrm{intra}}^{(k)} &= J_1 \sum_{j=1,2,3} \bm{S}_{1, j}^{(k)} \cdot \bm{S}_{2, j}^{(k)} +J_2 \sum_{i=1,2} \sum_{j=1,2} \bm{S}_{i, j}^{(k)} \cdot \bm{S}_{i, j+1}^{(k)}\notag\\
&+J_3 \sum_{j=1,2} \sum_{a=0,1} \bm{S}_{1,j+a}^{(k)} \cdot \bm{S}_{2, j+1-a}^{(k)}, \label{eq:hc}
\end{align}
and
\begin{equation}
\mathcal{H}_{\mrm{inter}}^{(k:k+1)} = J_{\mrm{inter}}\sum_{i=1,2} \bm{S}_{i,3}^{(k)} \cdot \bm{S}_{i,1}^{(k+1)} \label{eq:hi}
\end{equation}
where $\bm{S}_{i,j}^{(k)}$ is the  $S=\frac{1}{2}$ spin operator on the $(i,j)$ site of $k$-th cluster.
Here, the magnitude of inter-cluster interaction $J_{\mrm{inter}}$ is smaller than the intra-cluster interaction $J_{\mrm{intra}}=\sqrt{J_1^2+J_2^2+J_3^2}$.
If the rung interaction $J_1$ is FM (or weakly AFM), the ground state of the cluster can be a triplet, which is discussed below. %

\begin{figure}[t]
\begin{center}
\includegraphics[keepaspectratio,width=8.6cm]{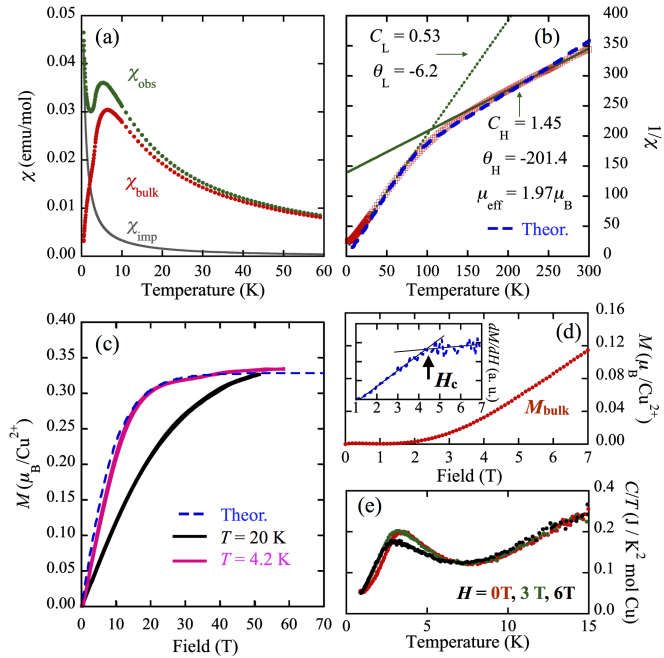}
\caption{(a) Temperature dependence of the magnetic susceptibility $\chi_{\rm bulk}$ (filled red circles) of K$_2$Cu$_3$O(SO$_4$)$_3$ measured at 0.1~T. 
The intrinsic susceptibility $\chi_{\rm bulk}$ is obtained by subtracting Pascal's diamagnetic contribution $\chi_{\rm dia}$ and an estimated contribution of impurity $\chi_{\rm imp}$  (gray solid line) from the experimental data $\chi_{\rm obs}$ (filled green circles). 
(b) The inverse susceptibility $1/\chi_{\rm bulk}$ (open red squares). 
The solid and broken green lines denote the fitting curves by the Curie-Weiss law.  
The blue dashed line represents the theoretical curve of the EST spin-cluster model (\ref{eq:hc}) with a weak FM interaction $J_1\cong-35$~K and strong AFM interactions $J_2=J_3\cong 125$~K.
(c) High-field magnetization at 4.2~K (pink solid line) and 20~K (black solid line). 
The blue dashed line denotes the theoretical magnetization curve obtained in the EST spin-cluster model with the same parameters in (b). 
(d) The magnetization of bulk $M_{\rm bulk}$ (filled red circles) and its field derivative $d$$M_{\rm bulk}$$/dH$ (blue broken line) measured at 0.52~K. The black solid lines are a guide to the eye.
(e) The temperature dependence of the total specific heat divided by temperature $C/T$ for 0 T, 3 T, and 5 T. % 
}
\label{MC}
\end{center}
\end{figure}

Figures~\ref{MC}(a), (b), and (d) present the magnetic susceptibility, the inverse susceptibility, and the magnetization curve of K$_2$Cu$_3$O(SO$_4$)$_3$. %
These are measured by using a SQUID magnetometer with a home-made $^3$He insert~\cite{kawae}. %
The intrinsic susceptibility $\chi_{\rm bulk}(T)$ is obtained by subtracting Pascal's diamagnetic contribution $\chi_{\rm dia}$~\cite{diamag} and the paramagnetic impurity moment $\chi_{\rm imp}=n_{\rm imp} C/(T-\theta_{\rm{CW}})$ from the experimental data $\chi_{\rm obs}$. 
Here, $C$ is the Curie constant for $S=\frac{1}{2}$ with the fixed $g$ factor as $g=2.0$, and $\theta_{\rm{CW}}$ is the Curie-Weiss temperature. %
As compared with the susceptibility at low temperatures $T\lesssim 3$~K in Fig.~\ref{MC}(a), we determine the impurity parameters as $n_{\rm imp}=0.032$, $C=1.12$, and $\theta_{\rm{CW}}=-0.3$~K, which indicate 3.2 percent of the magnetic moment originating from the impurity. %

Figure~\ref{MC}(b) shows the inverse susceptibility $1/\chi_{\rm bulk}$, which has a kink around 120~K.
The Curie constants and Weiss temperatures are estimated from the $1/\chi_{\rm bulk}$ by the Curie-Wiess law to be $C_{\rm L}= 0.53$ and $\theta$$_{\rm L}=-6.2$~K between 25 K and 55~K, and to be $C_{\rm H}=1.45$ and $\theta_{\rm H}=-201.4$~K between 200 K and 300~K. %
The $C_{\rm H}$ corresponds to an effective magnetic moment of 1.97 $\mu_{\rm B}$, which is consistent with a $S=\frac{1}{2}$ spin of Cu$^{2+}$. %
The Weiss temperature $\theta_{\rm H}=-201.4$~K indicates that strong AFM exchange interaction dominates in this system. %
The ratio of $C_{\rm L}$ and $C_{\rm H}$ is 0.38, which implies that only one-third of whole magnetic moments remains below 100~K. %
To clarify these features, we calculate the susceptibility by full diagonalization of the intra-cluster Hamiltonian (\ref{eq:hc}) without the inter-chain interaction $J_{\mrm{inter}}$.
As shown in Fig.~\ref{MC}(b), we obtain a good correspondence to experimental data with a weak FM interaction $J_1\cong-35$~K and strong AFM interactions $J_2=J_3\cong 125$~K, namely, the intra-cluster interaction $J_{\mrm{intra}}=180$~K.
Since these parameters satisfy the condition of Haldane state in ladder~\cite{Vekua}, the ground state of an EST spin-cluster is a triplet.

To confirm the triplet ground state of the EST spin-cluster, we measure the magnetization at high magnetic fields of up to 58 T. %
We can see a good coincidence between the theoretical and experimental data [see Fig.~\ref{MC}(c)], where the one-third plateau appears with small magnetic field as compared with the AFM interactions $J_2=J_3\cong 125$~K.
The good agreement strongly suggests that a spin-triplet state exists in K$_2$Cu$_3$O(SO$_4$)$_3$ in the high-temperature region. %
Therefore, the Haldane state is expected to be realised in the low temperature region.%

Next we consider the inter-cluster effect in K$_2$Cu$_3$O(SO$_4$)$_3$. %
The intrinsic magnetization $M_{\rm bulk}(H)$ is also obtained by subtracting the impurity moment from the experimental data by using the following equation $n_{\rm imp}gS\mu_{\rm B}[2\coth(gS\mu_{\rm B}H/k_{\rm B}T)-\coth(gS\mu_{\rm B}H/2k_{\rm B}T)]$ [Fig.~\ref{MC}(d)].
The $M_{\rm bulk}(H)$ shows $M=0$ plateau, which indicates the existence of a spin gap.
The critical field value $H_{\rm c}$ and the energy gap $\Delta_{MH}$ are related by the following equation $g$$\mu_B$$H_{\rm c}$ = $\Delta_{MH}$. 
The inset in Fig.~\ref{MC}(d) shows the field derivative of the magnetization curve $d$$M_{\rm bulk}$$/dH$. 
A continuous transition can be seen at $H_{\rm c}\approx 4.5$~T. 
Therefore, the excitation gap is obtained as $\Delta_{MH}\approx 6$~K. 

The existence of the spin gap in K$_2$Cu$_3$O(SO$_4$)$_3$ is also supported by magnetic susceptibility and specific heat measurements [Fig.~\ref{MC}(a) and (e)]. %
The $\chi_{\rm bulk}$ shows a broad maximum at around 4 K and decrease to zero with decreasing temperature. % 
 At same temperature, a Schottky-like peak in the heat capacity is observed. %
This peak corresponds to an energy gap $\Delta\approx 10$~K in two-level approximation. %
If the two levels originate from triplet and singlet of a local structure, the Schottky peak should show a field-sensitive gradual feature due to Zeeman splitting~\cite{FujihalaCdCu}. %
However, such a feature is not observed in the specific heat as shown in Fig.~\ref{MC}(e), and thus a local triplet-singlet gap does not explain the Schottky-like peak. %

\begin{figure}[t]
\begin{center}
\includegraphics[keepaspectratio,width=8.6cm]{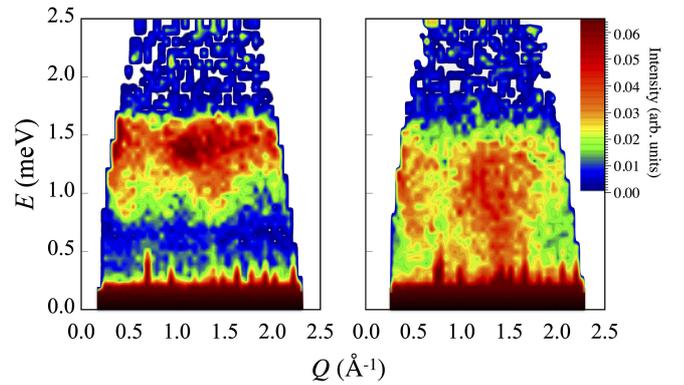}
\caption{The INS data by time-of-flight method in K$_2$Cu$_3$O(SO$_4$)$_3$, measured at (a) 1.5~K and (b) 4.0~K. 
}
\label{INS}
\end{center}
\end{figure}

To confirm the observation of the spin gap and to determine its magnitude, we performed an inelastic neutron scattering (INS) measurement. %
Data was acquired using the cold-neutron time-of-flight spectrometer PELCIAN. %
The INS data (Fig.~\ref{INS}) shows a gapped spectrum at 1.5 K; this spin gap has closed in the data collected at 4 K. %
Inspection of the elastic line revealed no additional Bragg reflections, showing that there is no magnetic long range order. %
The lower bound of the gap corresponds to a $\Delta_{\rm{INS}}$ = 0.61(3)~meV = 7.1~K. %
This is consistent with that determined by magnetization. %
The observation of dispersive modes in the INS data is consistent with K$_2$Cu$_3$O(SO$_4$)$_3$ being highly one-dimensional. %
A general feature of powder averaged data is that the zone boundary contributions are dominant. %
One consequence of this is that the observed dispersion minima are slightly off from the Brillouin zone centers $\pi$/$b$ and $3\pi$/$b$ to the high-$Q$ side. %
However it is reasonable to assume that this is the powder averaged spectrum where the excitations only disperse along one crystallographic direction~\cite{Stone,Nilsen}. %
We conclude from both experimental and theoretical results that the ground state of K$_2$Cu$_3$O(SO$_4$)$_3$ is a cluster based Haldane state. %

Finally, we discuss the cluster-based Haldane state from theoretical point of view, and estimate the inter-cluster interaction. 
The cluster-based $S=1$ spin operators in $k$th cluster $\bm{T}_{k}$ are analitically obtained by triplet ground states of the cluster~\cite{note1}.
Figure~\ref{theo}(a) shows a schematic configuration of the $T^z=1$ state of the cluster-based $S=1$ spin. 
The projection operator on the triplet states, $\mcl{P}^{(k)}$, gives us a relation between the edge spins $\bm{S}_{i,j}^{(k)}$ ($i=1,2$ and $j=1,3$) and the cluster-based spin: $\mcl{P}^{(k)}\bm{S}_{i,j}^{(k)}\mcl{P}^{(k)} = \frac{3}{8} \bm{T}_{k}$ for $i=1,2$ and $j=1,3$.
Therefore, we obtain the cluster-based Haldane chain as follows,
\begin{equation}
  \mathcal{H}_{\mrm{eff}}=\mcl{P}\mathcal{H}\mcl{P} =J_{\mrm{eff}} \sum_{k} \bm{T}_{k} \cdot \bm{T}_{k+1}+\mrm{const.},
\end{equation}
where the projection is given by $\mcl{P}=\prod_k\mcl{P}^{(k)}$ and the effective exchange energy $J_{\mrm{eff}}$ between cluster-based $S=1$ spins [see Fig.~\ref{theo}(b)] corresponds to $9J_{\mrm{inter}}/32$. %
In this effective model, the Haldane gap in the thermodynamical limit is obtained to be $\Delta_{\mrm{eff}}\sim 0.41 J_{\mrm{eff}} \cong 0.12 J_{\mrm{inter}}$~\cite{Haldane-gap}, when the inter-cluster interaction $J_{\mrm{inter}}$ is much smaller than the intra-cluster interaction $J_{\mrm{intra}}$.

Figure~\ref{theo}(c) shows the Haldane gap as a function of inverse system size 1/$N_{\mrm{EST}}$ for various $J_{\mrm{inter}}/J_{\mrm{intra}}=0.1,0.2,$ and $0.3$, which are numerically obtained by variational matrix-product state method~{\cite{schollwoch,note2} in the model Hamiltonian (\ref{eq:ham}) with the  intra-cluster interaction $J_{\mrm{intra}}$ = 180~K.
We can confirm that the Haldane gap $\Delta_{\mrm{calc}}\sim 2.2$~K in the thermodynamical limit $N_{\mrm{EST}}\to\infty$ for $J_{\mrm{inter}}=0.1J_{\mrm{intra}}=18$~K corresponds to $\Delta_{\mrm{eff}}$ obtained in the effective model, and therefore the numerical result verifies emergence of the cluster-based Haldane state for a small inter-cluster interaction.
By comparing with the gap obtained by INS experiment $\Delta_{\rm{INS}}\cong 7$~K, the inter-cluster interaction in K$_2$Cu$_3$O(SO$_4$)$_3$ is roughly estimated to be $J_{\mrm{inter}}\cong 0.2 J_{\mrm{intra}}\cong 36$~K. 
Since the estimated inter-cluster interaction is small enough as compared with the intra-cluster interaction, we conclude that the condition of the cluster-based Haldane state in K$_2$Cu$_3$O(SO$_4$)$_3$. 
%This calculation results indicate that the Haldane gap cannot be estimated by $0.12 J_{\mrm{inter}}$ even in the case of $J_{\mrm{inter}}/J_{\mrm{intra}}=0.2$.

Furthermore, we propose an extension of our model to general EST-SCCs where the number of tetrahedra $N_{\mrm{tetra}}$ in the EST cluster is adjustable.
According to the present study where $N_{\mrm{tetra}}=2$, the cluster-based Haldane state requires the ground state to be a triplet in the EST cluster.
If the exchange energy $|J_1|$ is smaller than AFM $J_2$ and $J_3$, the ground state is the same as the Haldane state in ladder~\cite{Vekua}.
In this case, the number of tetrahedra in the cluster is crucial in realizing triplet in the ground state.
Figure~\ref{theo}(d) shows the ground-state energy of singlet and triplet states as a function of $N_{\mrm{tetra}}$, where we can see an alternating behavior of singlet and triplet levels.
Consequently, we conclude that the cluster-based Haldane state is realized in the general EST-SCCs when the number of tetrahedra is $even$, i.e., $N_{\mrm{tetra}}=0$ (mod 2).

\begin{figure}[t]
\begin{center}
\includegraphics[keepaspectratio,width=8.6cm]{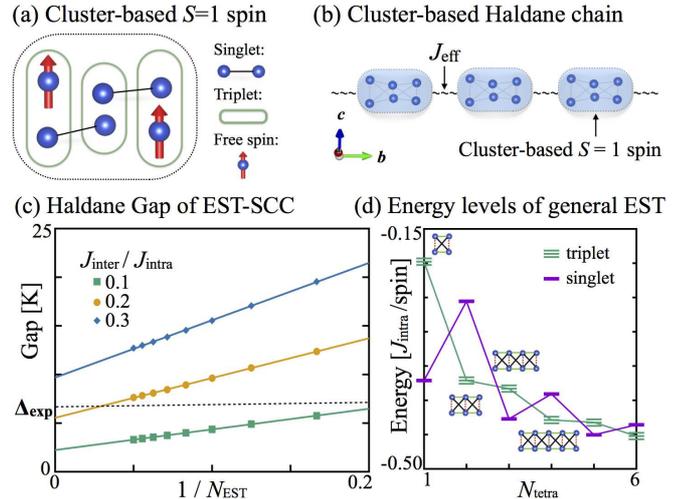}
	\caption{
(a) Schematic spin configuration of the $T^z=1$ state of cluster-based $S=1$ spin. This state is a Haldane state in a three-rung ladder and thus each rung composes a triplet (green oval) and anti-symmetrically connects with neighbor rungs (solid line). The free spin on edges is denoted by a red arrow. %
(b) Schematic effective model of cluster-based $S=1$ Haldane chain. Each cluster composes a cluster-based $S=1$ spin, and the effective exchange interaction originates from the inter-cluster interaction. %
(c) Extrapolation of the Haldane gaps as a function of inverse system size $N_{\mrm{EST}}$ for various $J_{\mrm{inter}}$. The Haldane gap is defined by the energy gap between $M=1$ and $2$ ground states. The black dashed line indicates the magnitude of the gap obtained by INS experiment. %
(d) Ground-state energies of singlet and triplet versus number of tetrahedra in a generalized EST spin-cluster. Here, we neglect inter-cluster interaction, and use the same intra-cluster super-exchange interactions as the estimated values for K$_2$Cu$_3$O(SO$_4$)$_3$ corresponding to the case of $N_{\mrm{tetra}}=2$.
}
\label{theo}
\end{center}
\end{figure}

In summary, we have found that Fedotovite K$_2$Cu$_3$O(SO$_4$)$_3$ has a unique arrangement of magnetic ions and a two-stage magnetic behaviors divided by a characteristic temperature $T\cong 4$~K.
The high-temperature experiments have given a good correspondence to theoretical thermodynamical values of an EST spin-cluster whose ground state is a triplet based on the Haldane state in a three-rung ladder.
In the lower-temperature region, the experimental data indicate the existence of a spin gap, which is expected by the cluster-based $S=1$ Haldane state.
Furthermore, we have theoretically obtained an effective Hamiltonian of a cluster-based $S=1$ Haldane chain, and estimated the inter-cluster interaction of K$_2$Cu$_3$O(SO$_4$)$_3$ by comparing the spin gap observed in the INS experiment.
Our study can propose an extension of general EST-SCCs exhibiting the cluster-based $S=1$ Haldane state, when the number of tetrahedra is $even$.
K$_2$Cu$_3$O(SO$_4$)$_3$ as well as general EST-SCCs is a promising candidate for qubits making use of Haldane states.~\cite{Qbit}.

Travel expenses for the inelastic neutron scattering experiments performed using PELICAN at ANSTO, Australia, were supported by General User Program for Neutron Scattering Experiments, Institute for Solid State Physics, The University of Tokyo (proposal no. 16900), at JRR-3, Japan Atomic Energy Agency, Tokai, Japan. Synchrotron powder XRD measurements were performed with the approval of the Photon Factory Program Advisory Committee (Proposal Nos.2015P001 and 2016G030).%


\begin{thebibliography}{99}\label{sec:TeXbooks}
\bibitem{TLL nature_Mat} I. A. Zaliznyak, Nat. Mater. {\bf 4}, 273 (2005).% 
\bibitem{I. Affleck} I. Affleck,  J. Phys.: Condens. Matter {\bf 1}, 3047 (1989). %
\bibitem{haldane's conjecture-1} F. D. M. Haldane, Phys. Rev. Lett. {\bf 50}, 1153 (1983).%
\bibitem{haldane's conjecture-2} F. D. M. Haldane, Phys. Lett. A {\bf 93}, 464 (1983).%
\bibitem{RiceLadder} T. M. Rice, S. Gopalan, and M. Sigrist, Europhys.Lett. {\bf 23}, 445 (1993).%
\bibitem{Matsuda2leg} T. Masuda, A. Zheludev, H. Manaka, L.-P. Regnault, J.-H. Chung, and Y. Qiu, Phys. Rev. Lett. {\bf 96}, 047210 (2006).%
\bibitem{Qbit} F. Meier, J. Levy, and D. Loss, Phys. Rev. Lett. {\bf 90}, 047901 (2003).
\bibitem{Vekua} T. Vekua and A. Honecker, Phys. Rev. B {\bf 73}, 214427 (2006).
\bibitem{mineralFed} L. P. Vergasova, S. K. Filatov, Y. K. Serafimova, and G. L. Starova, Doklady Acad. Nauk SSSR, {\bf 299}, 961 (1988). %
\bibitem{Fujio} F. Izumi and K. Momma, Solid State Phenom. {\bf 130}, 15 (2007).%
\bibitem{mineralFedstructure} G. L. Starova, S. K. Filatov, V. S. Fundamensky, and L.P. Vergasova, Mineral. Mag., {\bf 55}, 613 (1991). %
\bibitem{supple} See Supplemental Material at http://link.aps.org/ for the experimental details and crystal structure.
\bibitem{Misuno} Y. Mizuno, T. Tohyama, S. Maekawa, T. Osafune, N. Motoyama, H. Eisaki, and S. Uchida, Phys. Rev. B {\bf 57}, 5326 (1998). %
\bibitem{kawae} Y. Sato, S. Makiyama, Y. Sakamoto, T. Hasuo, Y. Inagaki, T. Fujiwara, H. S. Suzuki, K. Matsubayashi, Y. Uwatoko, and T. Kawae, Jpn. J. Appl. Phys. {\bf 52}, 106702 (2013). %
\bibitem{diamag} G. A. Bain, and J. F. Berry, J. Chem. Educ. {\bf 85}, 532 (2008). %
\bibitem{FujihalaCdCu} M. Fujihala, X. G. Zheng, H. Morodomi, T. Kawae, A. Matsuo, K. Kindo, and I. Watanabe, Phys. Rev. B {\bf 89}, 100401(R) (2014). 
\bibitem{Stone} M. B. Stone, G. Ehlers, and G. E. Granroth, Phys. Rev. B {\bf 88}, 104413 (2013). %
\bibitem{Nilsen}  G. J. Nilsen, A. Raja, A. A. Tsirlin, H. Mutka, D. Kasinathan, C. Ritter, and H. M. R$\o$nnow, New J. Phys. {\bf 17}, 113035 (2015).
\bibitem{note1} More explicitly, the cluster-based $S=1$ spin operators are given by $T_{k}^z = \sum_{\alpha=\pm,0} \alpha\, |\alpha^{(k)}\rag \lag \alpha^{(k)}|$ and $T_{k}^\pm = \sqrt{2} \lt(|\pm^{(k)}\rag \lag 0^{(k)}| +|0^{(k)}\rag \lag \pm^{(k)}|\rt)$ where $|\alpha^{(k)}\rag$ ($\alpha=\pm$ or 0) denotes the triplet ground state as the Haldane state in $k$th cluster. In this expression, the projection operator is also given by $\mcl{P}^{(k)}K=\sum_{\alpha=\pm,0} |\alpha^{(k)}\rag \lag \alpha^{(k)}|$.
\bibitem{Haldane-gap} S. R. White, Phys. Rev. Lett. {\bf 69}, 2863 (1992).
\bibitem{note2} This method is mathematically the same as well-known density-matrix renormalization-group method but has potential for extensive use.
\bibitem{schollwoch} U. Schollw\"och, Annals of Physics {\bf 326} 96 (2011).
\end{thebibliography}
\end{document}